\documentclass[lettersize,journal]{IEEEtran}
\usepackage{amsmath,amsfonts}
\usepackage{algorithmic}
\usepackage{algorithm}
\usepackage{array}
\usepackage[caption=false,font=normalsize,labelfont=sf,textfont=sf]{subfig}
\usepackage{textcomp}
\usepackage{stfloats}
\usepackage{url}
\usepackage{verbatim}
\usepackage{graphicx}
\usepackage{cite}
\usepackage{xcolor}
\usepackage{amssymb}

\usepackage{amsthm}
\newtheorem{proposition}{Proposition}

\newcommand{\KL}{D_{\mathrm{KL}}} 
\newcommand{\rev}[1]{\textcolor{black}{#1}}

\begin{document}

\title{Symbol Distributions in Semantic Communications: A Source-Channel Equilibrium Perspective}


\author{Hanju~Yoo,~\IEEEmembership{Graduate Student Member,~IEEE,}
    Dongha~Choi,~\IEEEmembership{Graduate Student Member,~IEEE,}
    Songkuk~Kim,~\IEEEmembership{Member,~IEEE,}  
    Chan-Byoung~Chae,~\IEEEmembership{Fellow,~IEEE,} and Robert W. Heath, Jr.,~\IEEEmembership{Fellow,~IEEE,}
    \thanks{H. Yoo, D. Choi, S. Kim, and C.-B. Chae are with the School of Integrated Technology, Yonsei University, Seoul 03722, South Korea (e-mail: \{hanju.yoo, ellijah1030, songkuk, cbchae\}@yonsei.ac.kr).}
    \thanks{R. W. Heath, Jr. is with the Department of Electrical and Computer Engineering, University of California San Diego, La Jolla, CA, 92093 USA (email: rwheathjr@ucsd.edu).}%
\thanks{Manuscript received Nov. XX, 2025; revised Dec. XX, 2025.}}

\markboth{submitted to IEEE Trans. on Communications, Nov.~2025}%
{Yoo \MakeLowercase{\textit{et al.}}: Symbol Distributions in Semantic Communications: A Source-Channel Equilibrium Perspective}

\IEEEpubid{0000-0000/00\$00.00~\copyright~2025 IEEE}

\maketitle

\begin{abstract}
Semantic communication systems often use end-to-end neural networks to map input data into continuous symbols. These symbols, which are essentially neural network features, have fixed dimensions and often exhibit heavy-tailed distributions.
However, the mechanism behind this distributional shape remains underexplored due to the end-to-end nature of encoder training, hindering systematic analysis and design.
In this paper, we propose a parametric model for semantic symbol distributions. 
\rev{We model end-to-end training as inducing two coupled pressures on the symbol distribution: a \emph{source pressure} that favors power allocation minimizing the average description cost, and a \emph{channel pressure} that favors distributions with higher channel utilization.
Under surrogate objectives that capture these effects, we obtain a Student's $t$-distribution as a model for the semantic symbols.
Experiments on image-based semantic systems show that the model closely predicts how the shape parameter varies with (i) explicit symbol rate control and (ii) dataset entropy variability.
Furthermore, enforcing a target symbol distribution via regularization (e.g., a Gaussian prior) improves training convergence, which is consistent with our hypothesis.}

\end{abstract}


\section{Introduction}

\IEEEPARstart{S}{emantic} communications have emerged as a deep learning–based approach that jointly optimizes the entire communication process~\cite{BeyondBits}. Instead of preserving bit-level fidelity, semantic systems focus on transmitting task-relevant information, enabling efficient communication for tasks such as object detection, image classification, and data reconstruction~\cite{yoo2022real,bourtsoulatze2019deep,xie2021deep,weng2021semantic,tung2021deepwive,xu2021wireless}.

Analog semantic communication systems such as deep joint source–channel coding (DeepJSCC) consist of an end-to-end neural network encoder and decoder. The encoder maps input data into low-dimensional, continuous-valued symbols, contrasting with conventional digital modulation based on discrete constellations (e.g., QAM).
Given a noisy channel and finite power constraints, the encoder must pursue two goals: (i) representing the source with as little redundancy as possible (source-coding efficiency) and (ii) shaping the transmitted symbols to carry information reliably through the channel (channel-coding efficiency)\footnote{In this work, we use the term `channel coding' in its broad information-theoretic sense, including all mechanisms aimed at maximizing the mutual information across the channel.}. 
In conventional systems, these goals can be handled by separate modules and variable-length bitstrings. In contrast, most semantic encoders emit a \emph{fixed-length} vector of continuous symbols~\cite{yoo2022real,bourtsoulatze2019deep,xie2021deep,weng2021semantic}, so both goals must be met primarily by shaping the same transmitted-symbol distribution under a power constraint. 
This coupling creates an inherent trade-off, yielding symbol distributions that are neither purely source- nor purely channel-coding oriented.
The resulting symbol distributions of these learned systems remain largely unexplained, as end-to-end training with task-specific loss functions obscures the mechanisms shaping the symbols. This hinders systematic analysis and principled design of practical semantic systems~\cite{yoo2025bridging}.

\rev{In this paper, we develop a simple parametric framework for the symbol distributions learned by end-to-end semantic encoders, enabling more principled analysis and design of practical semantic transceivers.
Our goal is not to propose a new method for maximizing semantic reconstruction performance, but rather to model the learned symbol statistics that emerge in such systems.
We focus on fixed-length analog semantic systems, where the encoder cannot separately optimize source compression and transmission reliability.
Instead, both effects must be expressed through the same fixed-length continuous symbol vector under an average-power constraint, i.e., through the shape of symbol distribution.} \IEEEpubidadjcol

\rev{We formalize this coupling by interpreting end-to-end training as inducing two concurrent pressures on the symbol distribution:}
\begin{itemize}
\item \textbf{Channel pressure.} The symbol distribution is driven to use the channel efficiently. In AWGN, mutual-information–driven channel utilization is naturally related to entropy maximization in the Shannon asymptotic regime.
\item \textbf{Source pressure.} The symbol distribution is driven to pack more source information into a fixed-length symbol vector under an average-power constraint. This is promoted by a frequency-energy assignment in which low-energy symbols represent frequent features while high-energy symbols represent rare ones, effectively realizing an implicit variable-length description and inducing heavy-tailed amplitudes.
\end{itemize}


\rev{To capture these competing pressures in a tractable form, we introduce surrogate objectives for description efficiency and channel utilization, leading to a parametric Student's t model for the learned symbol distribution. We then empirically validate this model using trained DeepJSCC systems.}

\begin{figure*}[htbp]
  \centering
  \includegraphics[width=0.9\textwidth]{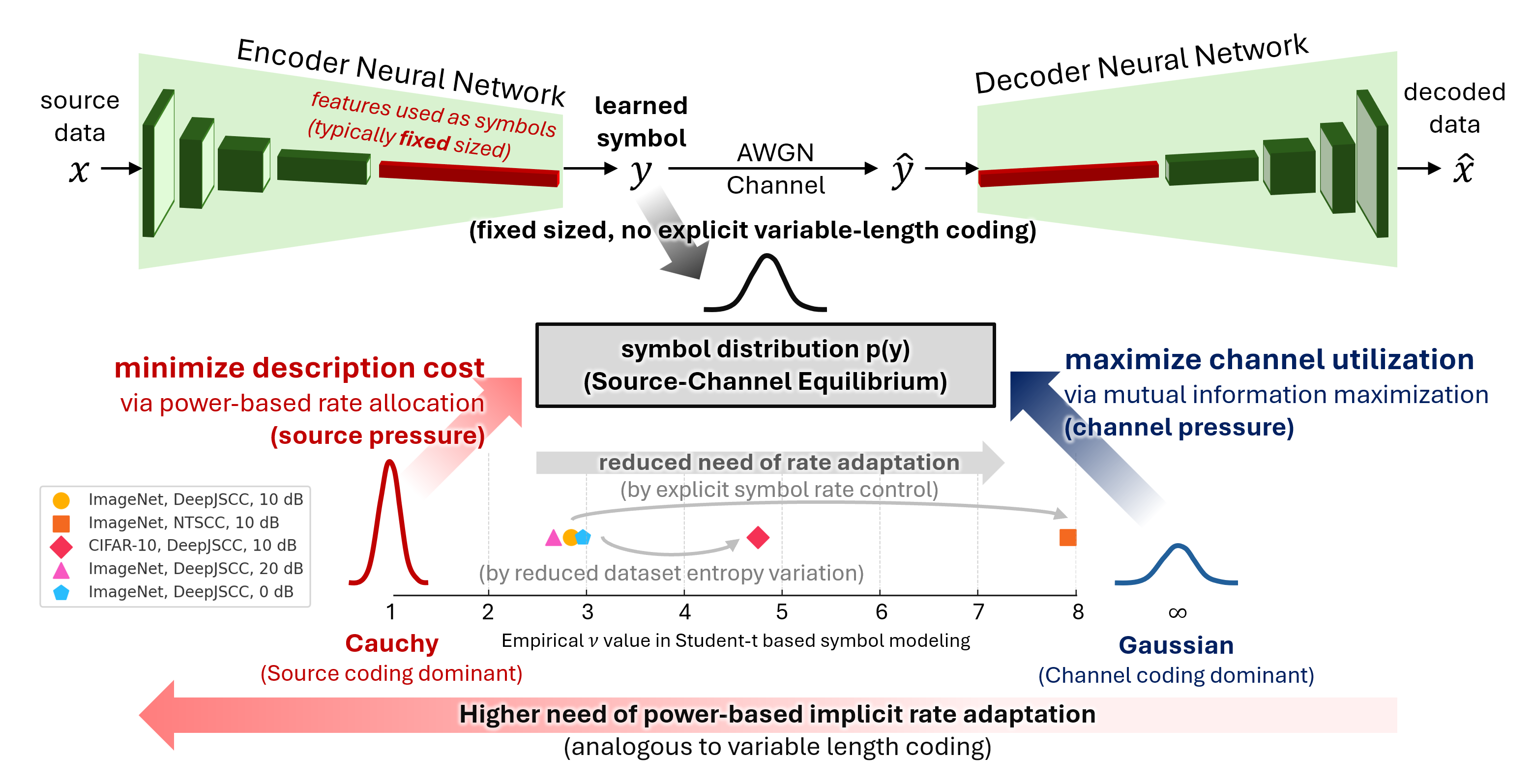}
\caption{Conceptual diagram of fixed-length semantic communication over an AWGN channel. 
End-to-end learning shapes the symbol distribution $p(Y)$ by balancing source pressure (description efficiency) and channel pressure (mutual-information maximization), yielding behaviors ranging from heavy-tailed (Cauchy-like) to Gaussian-like distributions. Empirical evidence for this trend is reported in Section~\ref{sec:empirical_valid}.}
  \label{fig:semantic_arch}
\end{figure*}

The main contributions of this paper are as follows:
\begin{itemize}
    \item \textbf{Parametric modeling of semantic symbol distributions.} 
    We model semantic encoders as balancing two competing objectives: maximizing mutual information over the channel (channel pressure) and reducing the effective bitlength of transmitted symbols via power allocation (source pressure). By formulating this trade-off through tractable surrogate objectives, we show that the resulting symbol distribution follows a Student's $t$-distribution.

    \item \textbf{Empirical validation of the proposed model.} 
    Extensive experiments on image-based semantic communication systems demonstrate that learned symbol distributions closely match the proposed Student's $t$-model. Moreover, the observed variations in the shape parameter are consistent with the model predictions, particularly with respect to (i) the use of variable-length coding mechanisms and (ii) the entropy characteristics of the dataset.

    \color{black}
    \item \textbf{Distribution-aware regularization as empirical support for the framework.} 
    Building on the proposed model, we introduce a regularization loss that explicitly guides the encoder’s empirical symbol distribution toward a target prior. Experimental results show that this approach can improve training convergence and stability, providing practical evidence that symbol-distribution shaping influences end-to-end learning behavior.
    \color{black}
\end{itemize}

\section{Prior Work}
\label{sec:prior_work}

Deep learning-based end-to-end optimization has been successfully applied to semantic communication systems.
The foundational work~\cite{bourtsoulatze2019deep} introduced DeepJSCC, which directly maps images to in-phase and quadrature-phase (I/Q) symbols via neural networks. This approach demonstrates superior reconstruction quality compared to conventional systems and ensures graceful performance degradation, maintaining acceptable image quality even in low SNR regimes without the need for retransmission.
Subsequent research validates these semantic systems through hardware prototypes~\cite{yoo2022real, yoo2023role, ding2024adaptive, liu2022real, yoo2025bridging}, demonstrating robustness in real-world wireless environments.
However, the black-box nature of the learned encoder makes it difficult to incorporate practical hardware constraints in a principled way~\cite{yoo2025bridging}, which can lead to performance degradations in over-the-air deployments~\cite{yoo2022real, yoo2023role, ding2024adaptive, liu2022real, yoo2025bridging}.

\rev{Classical joint source-channel coding (JSCC) theory also provides a useful point of reference for our work~\cite{gastpar2003source,gastpar2003code}. In particular, source-channel matching results characterize efficient communication through coupled source-side and channel-side information-theoretic structure~\cite{gastpar2003code}. On the channel side, the optimality conditions can be written through a Kullback-Leibler (KL)-divergence quantity relative to the induced output distribution; in the Gaussian/AWGN case, this reduces to the familiar quadratic power-cost viewpoint~\cite{gastpar2003code}. On the source side, optimal communication can be interpreted as requiring a match between the source-side information requirement and the channel-side information budget at the operating point~\cite{gastpar2003code,gastpar2003source}. Our work is consistent with this matching perspective, but addresses a different problem: rather than casting the problem as exact matching, we view the practical learned semantic system through an optimization trade-off between source-side and channel-side objectives, and use tractable surrogate terms analogous to the corresponding classical matching principles to model the continuous symbol distribution learned under fixed-length analog transmission.}

\rev{The DeepJSCC architecture is closely related in spirit to learned image compression~\cite{balle2016end,balle2018hyperprior,cheng2020learned}, where neural transforms are trained to optimize a rate-distortion trade-off.}
\rev{In learned compression, the rate is typically the entropy of a quantized latent representation, which is converted into a variable-length bitstream via entropy coding. During training, this rate is estimated using learned latent priors~\cite{balle2016end,balle2018hyperprior,cheng2020learned}, yielding a Lagrangian loss that combines distortion and estimated entropy.}
\rev{Our analysis is also inspired by rate-distortion and information-theoretic concepts, but serves a different purpose.
Rather than defining a training loss as in learned compression, we use a Lagrangian formulation as an analysis tool to explain and derive the distributional form of the continuous semantic symbols produced by an end-to-end trained encoder.
Moreover, unlike learned compression, which relies on quantization and variable-length bitstreams, our setting is fixed-length and continuous-valued: rate adaptation cannot be realized by changing codeword lengths, and must instead be reflected in the symbol distribution itself.}

Recent theoretical studies have explored analysis frameworks such as semantic entropy or semantic capacity~\cite{niu2024mathematical, shao2024theory}.
While conceptually valuable, these high-level frameworks do not address the specific engineering problem of modeling the physical-layer symbol distributions generated by a DeepJSCC encoder.
The most closely related works leverage the information bottleneck (IB) principle or Variational Autoencoders (VAEs)~\cite{xie2023robust, feng2024variational}.
For instance, \cite{xie2023robust} adopts an information-theoretic perspective leveraging symbol entropy.
However, this approach inherently treats channel noise as part of the VAE's stochastic sampling mechanism, focusing the analysis on the post-noise symbols.
The specific distribution generated by the encoder before noise addition, which reflects the encoder's inherent power allocation strategy, remains underexplored.
Our paper addresses this gap by providing the first symbol distribution model and empirical validation for the shape of the pre-noise semantic symbol distribution.

\rev{Recent vector quantized (VQ)-VAE-based regularization approaches are also related in motivation, since mutual-information or KL-based objectives can be interpreted as encouraging better utilization of a fixed communication interface. In particular, the mutual-information-regularized VQ-VAE framework in~\cite{chen2026precoding} discourages codeword collapse and promotes more uniform usage of a discrete codebook under a fixed feedback budget. This is closely related in spirit to our channel-utilization pressure. However, the modeling target is different: the cited work regularizes a discrete latent representation, whereas we study the continuous pre-noise transmitted symbol distribution itself and explain its shape through the balance between channel utilization and source-side description efficiency.}

Another relevant line of research is probabilistic constellation shaping (PCS), which optimizes symbol distributions under power constraints to approach channel capacity, including both classical Maxwell-Boltzmann shaping and recent deep learning-based geometric/probabilistic shaping methods~\cite{kschischang1993optimal,bocherer2015bandwidth,forney1984efficient,stark2019joint,aref2022end}.
\rev{Analogously, our semantic encoder can be interpreted as performing an implicit form of probabilistic shaping to maximize transmission efficiency under average power constraints.
However, while existing PCS methods remain in discrete signaling for channel throughput maximization, our semantic encoder operates in a continuous symbol space and optimizes a joint source-channel objective.
Our analysis reveals that this dual source-channel pressure yields a Student's $t$ distribution, which contrasts with the channel-centric Maxwell-Boltzmann distribution commonly encountered in probabilistic constellation shaping.} 

\section{End-to-End Trained Semantic Communication Systems}
This section reviews the standard end-to-end architecture of neural semantic communication systems and introduces the notation used in our analysis.
Throughout the section, scalar random variables are denoted by capitals (e.g., $X,Y$),
and their realizations by lowercase letters (bold for vectors), e.g., $x,y,\mathbf{x},\mathbf{y}$.

\label{sec:e2e_semantic_systems}
\subsection{Typical Architecture of Semantic Communication Systems}

Fig.~\ref{fig:semantic_arch} depicts a typical architecture for a semantic communication system which comprises three main components: an encoder, an AWGN channel, and a decoder.
The encoder, usually implemented as a deep neural network, maps an input data vector $\mathbf{x}$ (e.g., an image or text) to a vector of complex-valued symbols $\mathbf{y}$ that capture the essential semantic information:
\begin{equation}
\mathbf{y} = f_{\text{enc}}(\mathbf{x}),
\end{equation}
where $f_{\text{enc}}(\cdot)$ denotes the encoding function realized by the network.

To enforce the power constraint during neural network training, the encoder output is normalized based on the empirical second moment over the training batch.
Specifically, to approximate the unit average power condition (i.e., $\mathbb{E}[\|\mathbf{Y}\|^2]=1$), the symbol vector is scaled such that its average squared norm over the batch equals one:
\begin{equation}
\tilde{\mathbf{y}} = \frac{\mathbf{y}}{\sqrt{\frac{1}{B}\sum_{b=1}^{B} \left\| \mathbf{y}^{(b)} \right\|^2}},
\label{eq:power_norm}
\end{equation}
where $B$ is the batch size, $\mathbf{y}^{(b)}$ denotes the symbol vector realization for the $b$-th sample in the batch, and $\tilde{\mathbf{y}}$ is the normalized symbol vector.
\rev{The batch-wise power normalization is part of the standard DeepJSCC transmitter and therefore part of the learned system whose output distribution we analyze. For distribution fitting, we collect the normalized transmitted symbols over the evaluation set and apply only a single global variance normalization to fix the scale. This analysis-time rescaling changes the scale of the analyzed marginal but not its distributional shape, which is consistent with our use of a variance-normalized Student's $t$ model.}

The normalized symbol vector $\tilde{\mathbf{y}}$ is then transmitted through a differentiable channel layer that models the physical communication environment, typically incorporating noise and attenuation.
A common choice is the AWGN channel, modeled as:
\begin{equation}
\hat{\mathbf{y}} = \tilde{\mathbf{y}} + \mathbf{n}, \quad \text{where } \mathbf{n} \sim \mathcal{N}(\mathbf{0}, \sigma^2 \mathbf{I}).
\end{equation}
Here, $\hat{\mathbf{y}}$ represents the received symbols, $\sigma^2$ denotes the noise variance determined by the signal-to-noise ratio (SNR), and $\mathbf{I}$ is the identity matrix of appropriate dimensions.

The decoder, also implemented as a deep neural network, processes the received symbol vector $\hat{\mathbf{y}}$ to perform a specific task, such as reconstructing the original input:
\begin{equation}
\hat{\mathbf{x}} = f_{\text{dec}}(\hat{\mathbf{y}}).
\end{equation}
During training, a loss function is computed based on the decoder's output $\hat{\mathbf{x}}$. For reconstruction tasks, the loss is typically defined as the mean squared error (MSE) between the original input $\mathbf{x}$ and the reconstructed output $\hat{\mathbf{x}}$:
\begin{equation}
\ell = \frac{1}{B}\sum_{b=1}^{B} \left\| \mathbf{x}^{(b)} - \hat{\mathbf{x}}^{(b)} \right\|^2.
\end{equation}
We employ the MSE loss here for simplicity, but the core system architecture and our subsequent analysis are generalizable to other differentiable distortion metrics (e.g., perceptual or task-specific losses).
The parameters of both the encoder and decoder are updated via backpropagation to minimize $\ell$.

\begin{figure*}[htbp]
  \centering
  \includegraphics[width=\textwidth]{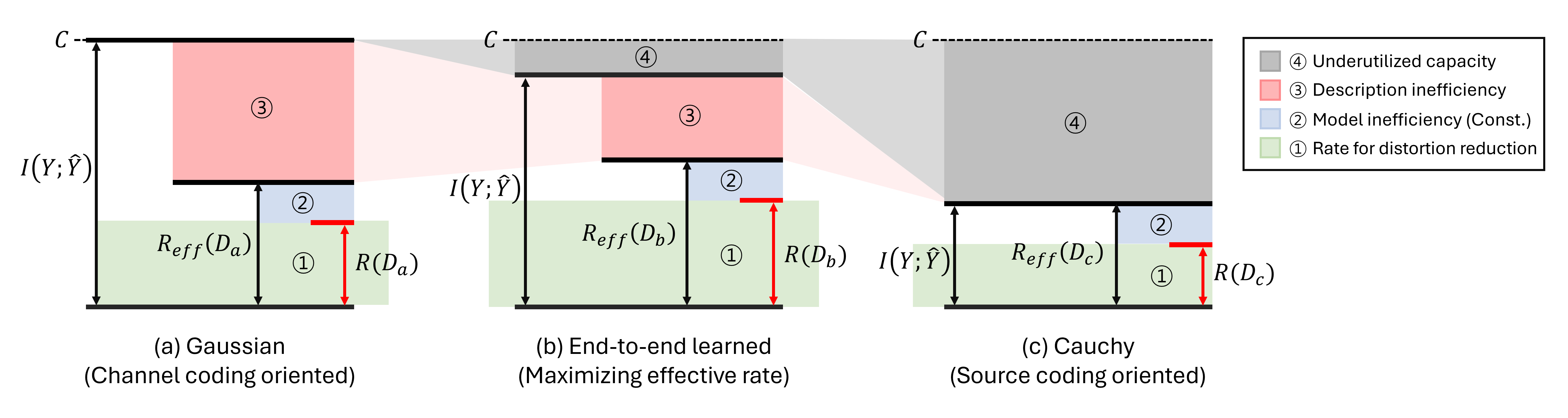}
    \caption{Rate decomposition for end-to-end semantic communication over AWGN at a fixed training SNR (with capacity $C$). 
    Different symbol distributions trade off underutilized capacity (4) and description inefficiency (3), yielding different operating distortions $D_a,D_b,D_c$.}
    \label{fig:entropy_analysis}
\end{figure*}

\color{black}
\subsection{Connection Between End-to-End Distortion and Source-Channel Coding Objectives}
\label{sec:insights_from_info_theory}

\rev{A semantic communication system is trained end-to-end to minimize a distortion $\ell$ between the source $X$ and the reconstruction $\hat X$. Although the training objective is task-level, it can be interpreted as inducing two coupled pressures on the symbol distribution $p(Y)$: the channel symbols should carry enough information to reduce distortion, but should not use substantially more channel information than is effectively required. We formalize this link through the following information-theoretic decomposition.}

\subsubsection{Distortion minimization induces an effective rate demand}
\rev{An ideal RD-optimal system would require only the Shannon rate-distortion limit $R(D)$ to achieve distortion level $D$, where}
\begin{equation}
\rev{R(D) \triangleq \inf_{p(\hat x|x):\, \mathbb{E}[d(X,\hat X)]\le D} I(X;\hat X).}
\label{eq:RD_def}
\end{equation}
From a distortion-rate viewpoint, achieving a smaller distortion requires a larger task-relevant information budget.
\rev{Accordingly, when end-to-end training pushes the system toward a lower distortion level, it also pushes the system toward an operating point associated with a larger $R(D)$.}

An end-to-end neural encoder/decoder pair is generally not RD-optimal, due to finite model capacity, finite blocklength effects, and optimization mismatch.
\rev{We capture all such nonidealities by defining an effective required rate for achieving distortion $D$,}
\begin{equation}
\rev{R_{\mathrm{eff}}(D) \triangleq R(D) + \delta_{\mathrm{model}},}
\qquad \delta_{\mathrm{model}}\ge 0,
\label{eq:Reff_def}
\end{equation}
where $\delta_{\mathrm{model}}$ aggregates architectural and training inefficiencies.
\rev{In our analysis of symbol shaping, we simply assume that $\delta_{\mathrm{model}}$ is a constant for a given architecture and training setup, and therefore independent of the symbol distribution $p(Y)$. This assumption reflects our intent to treat the effects of finite neural-network capacity and optimization mismatch as an aggregate non-ideality term, rather than to model their detailed dependence on the learned symbol distribution.}

\subsubsection{Symbol-generation stage and the channel information budget}
The encoder maps the input source $X$ to channel symbols $Y$, which pass through an AWGN channel
\begin{equation}
\hat Y = Y + N,\qquad N\sim\mathcal{N}(0,\sigma^2),
\end{equation}
with an average power constraint on $Y$.
The information-carrying capability of this symbol pipe is quantified by $I(Y;\hat Y)$ and is upper bounded by the channel capacity $C$ at the training SNR:
\begin{equation}
I(Y;\hat Y)\le C,
\label{eq:MI_upper_bound}
\end{equation}
where the capacity is defined as
\begin{equation}
C \triangleq \sup_{p(y):\,\mathbb{E}[|Y|^2]\le 1} I(Y;\hat Y).
\label{eq:capacity_def}
\end{equation}
\rev{To achieve a distortion level $D$, the channel must provide at least the effective rate required by the learned system. We therefore assume the system operates in the feasible regime}
\begin{equation}
\rev{R_{\mathrm{eff}}(D) \;\lesssim\; I(Y;\hat Y)\;\le\; C.}
\label{eq:operating_regime}
\end{equation}
\rev{Conversely, for a given learned symbol distribution, the available channel information budget $I(Y;\hat Y)$ determines the achievable distortion operating point of the end-to-end system; as this operating point shifts, the corresponding effective rate demand $R_{\mathrm{eff}}(D)$ also changes through its dependence on $D$.}

Importantly, the source-side rate requirement should not be identified directly with $I(Y;\hat Y)$.
The quantity $R_{\mathrm{eff}}(D)$ refers to the task-relevant source information needed by the model to attain distortion $D$, whereas $I(Y;\hat Y)$ measures the total information budget supported by the symbol interface.
In practice, the source information must be mapped onto a fixed number of continuous channel symbols, and this message-to-symbol assignment generally incurs additional representation overhead.
This overhead becomes more pronounced when the source statistics and symbol statistics are mismatched, or when explicit variable-length entropy coding is unavailable.
\rev{Therefore, only a portion of the channel information budget, rather than the full $I(Y;\hat Y)$, is used to convey the task-relevant source information quantified by $R_{\mathrm{eff}}(D)$.}

\subsubsection{Two pressures on $p(Y)$}
Eq.~\eqref{eq:operating_regime} implies that learning shapes $p(Y)$ through two opposing pressures:

\paragraph{Pressure 1: Channel utilization}
Any gap to capacity,
\begin{equation}
\delta_{\mathrm{under}} \triangleq C - I(Y;\hat Y)\ge 0,
\label{eq:delta_under}
\end{equation}
represents underutilized capacity (region~\textcircled{4} in Fig.~\ref{fig:entropy_analysis}).
Reducing $\delta_{\mathrm{under}}$ encourages larger $I(Y;\hat Y)$.
In the Shannon asymptotic setting for AWGN, this is closely linked to entropy maximization since the noise entropy is fixed. Motivated by this, we later model the channel pressure using a tractable entropy-based surrogate.

\paragraph{Pressure 2: Description efficiency}
At the same time, using a channel rate far above what is effectively required for the target distortion is wasteful.
We define the description inefficiency as
\begin{equation}
\delta_{\mathrm{desc}} \triangleq I(Y;\hat Y) - R_{\mathrm{eff}}(D)\ge 0,
\label{eq:delta_desc}
\end{equation}
corresponding to region~\textcircled{3} in Fig.~\ref{fig:entropy_analysis}.
Intuitively, $\delta_{\mathrm{desc}}$ is the ``excess pipe width'' consumed by the symbol distribution beyond what the task effectively demands at distortion $D$.
In classical source coding and learned image compression, this excess is reduced by variable-length bit allocation enabled by probability modeling and entropy coding.
In contrast, DeepJSCC-style semantic systems typically transmit a fixed number of continuous symbols per sample, so explicit variable-length coding is unavailable.
Consequently, rate adaptation must be realized implicitly within the fixed-length symbol space.
A natural mechanism is symbol-energy allocation: frequent features are transmitted with low energy, while rare features are assigned higher energy, playing a role analogous to variable-length coding.
This motivates a source-centric objective that penalizes the average per-symbol ``payload cost'' induced by the symbol amplitudes.

\subsubsection{Resulting equilibrium}
\rev{In summary, at a fixed training SNR, the neural network shapes $p(Y)$ by trading off channel utilization and description efficiency. As illustrated in Fig.~\ref{fig:entropy_analysis}, the balanced regime can achieve a smaller distortion than either extreme under the same capacity constraint. In the next section, we formalize these pressures using tractable surrogate objectives and show that the resulting maximum-entropy solution forms a Student's $t$ family.}

\color{black}
\section{Symbol Distributions for Semantic Communications}
\label{sec:optimal_distributions}

Building on the previous section, we view symbol shaping as the choice of an input distribution $p(y)$ under two competing pressures:
(i) \emph{channel utilization} (reducing the underutilized capacity $\delta_{\mathrm{under}}$) and
(ii) \emph{description efficiency} (reducing the overhead $\delta_{\mathrm{desc}}$ beyond the rate required at distortion $D$).

Directly optimizing $\delta_{\mathrm{under}}$ or $\delta_{\mathrm{desc}}$ over the empirical symbol distribution is intractable, so we introduce tractable surrogates and optimize a weighted composite objective:
\begin{equation}
\underset{p(y)}{\text{minimize}} \quad 
\mathcal{L}_{\mathrm{desc}}(p) + \lambda\,\mathcal{L}_{\mathrm{util}}(p),
\label{eq:weighted_objective_conceptual}
\end{equation}
where $\mathcal{L}_{\mathrm{util}}$ promotes channel utilization and $\mathcal{L}_{\mathrm{desc}}$ promotes description efficiency.
The weight $\lambda$ controls the trade-off between the two terms.
In the following subsections, we specify the surrogate objectives used to model description efficiency and channel utilization.

\subsection{Channel-Utilization Surrogate}
\label{subsec:channel_surrogate}
In the Shannon asymptotic setting, channel utilization can be interpreted as reducing the underutilized capacity
$\delta_{\mathrm{under}}\triangleq C-I(Y;\hat Y)$.
At a fixed training SNR, $C$ is constant; thus minimizing $\delta_{\mathrm{under}}$ is equivalent to maximizing $I(Y;\hat Y)$.
For an AWGN channel $\hat Y=Y+N$ with $N\sim\mathcal{N}(0,\sigma^2)$,
\begin{equation}
I(Y;\hat Y)=h(\hat Y)-h(N),
\end{equation}
where $h(N)$ is fixed.
\rev{While $I(Y;\hat Y)$ depends on $h(Y+N)$, the entropy power inequality implies that increasing $h(Y)$ tends to increase $h(Y+N)$ for fixed noise $N$, and Gaussian inputs maximize both $h(Y)$ under a power constraint and $I(Y;\hat Y)$ on AWGN. Motivated by this extremal property, we adopt $-h(Y)$ as a tractable distribution-level surrogate and define}
\begin{equation}
\mathcal{L}_{\mathrm{util}} \triangleq -h(Y),
\label{eq:channel_objective}
\end{equation}
\rev{so that minimizing $\mathcal{L}_{\mathrm{util}}$ encourages high-entropy input distributions.}\footnote{\rev{We use entropy maximization as a tractable, distribution-level surrogate, rather than as claims of finite-blocklength optimality.}}

\subsection{Description-Efficiency Surrogate via an Energy-Payload Model}
\label{subsec:source_surrogate}
In classical entropy coding, the optimal variable-length code assigns a codelength
$\ell(y)\approx -\log p(y)$, so that common symbols receive short descriptions and rare symbols receive long ones.
In contrast, our setting uses a fixed number of channel symbols per sample, so we cannot explicitly vary codeword lengths in bits.
If the system still needs per-sample (or per-feature) rate adaptation, a natural way is to realize it implicitly through symbol energy:
frequent features are transmitted with low energy (low ``payload'') and rare features with higher energy (high ``payload'').
This raises the key modeling question: \emph{how should instantaneous payload be related to symbol energy?}
To make this mechanism analyzable, we introduce an energy-payload surrogate $\ell_{\mathrm{pay}}(y)$ that maps a symbol amplitude to an effective number of reliably distinguishable states (or bits) under AWGN.

\begin{figure}[t]
  \centering
  \includegraphics[width=0.6\columnwidth]{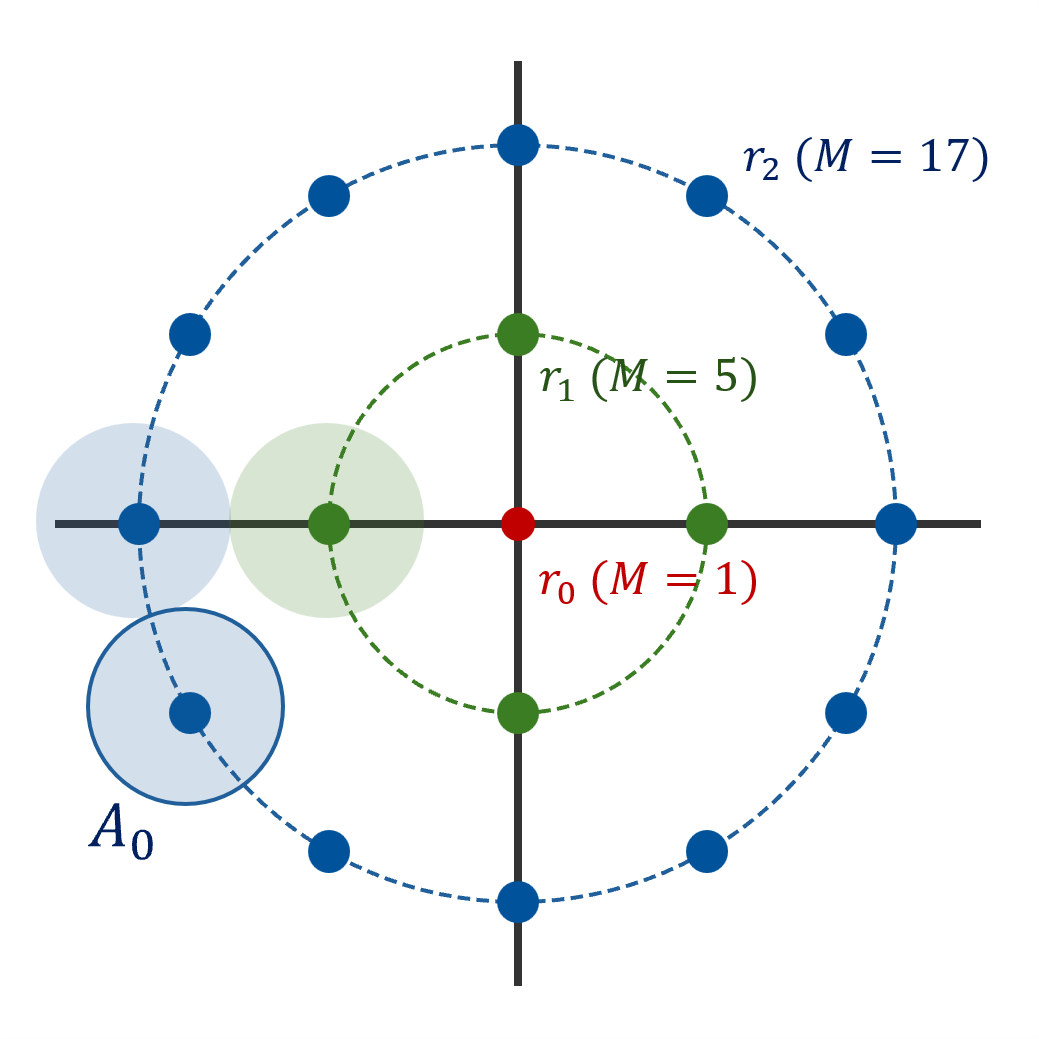}
  \caption{APSK-inspired packing view: the number of reliably distinguishable points within radius $r$ scales with area, hence $M(r)\propto r^2$.}
  \label{fig:apsk}
\end{figure}

\paragraph{APSK-inspired bit-to-power relation (heuristic)}
Consider a dense constellation on the I/Q plane designed to maintain a target error probability under AWGN with variance $\sigma^2$.
If reliable discrimination requires a minimum Euclidean separation $d_0$, then each decision region has an area on the order of
\begin{equation}
A_0 \;\asymp\; d_0^2.
\end{equation}
where $\asymp$ denotes equality up to a constant scaling factor.
Under approximately uniform packing, the number of distinguishable points inside a disk of radius $r$ scales with its area:
\begin{equation}
M(r) \;\approx\; 1 + \kappa\,\frac{\pi r^2}{A_0},
\label{eq:Mr_affine}
\end{equation}
where the additive $1$ enforces $M(0)=1$ (zero payload at the origin) and $\kappa\in(0,1]$ captures packing efficiency.
The corresponding payload in nats is then
\begin{equation}
b(r) \;\triangleq\; \ln M(r)
\;=\;
\ln\!\left(1+\kappa\,\frac{\pi r^2}{A_0}\right).
\label{eq:br_def}
\end{equation}

\paragraph{From geometry to energy}
We relate the geometric radius to symbol energy by identifying $r^2$ with the instantaneous squared magnitude of the transmitted symbol, i.e., $r^2 \propto |y|^2$ (for a real scalar component, $|y|^2=y^2$).
All constants in \eqref{eq:Mr_affine}-\eqref{eq:br_def}, including the packing factor $\kappa$, the region area scale $A_0$, and the dependence of the required separation $d_0$ on the operating noise level, are absorbed into a single nonnegative constant $\beta$.
This yields the per-symbol payload surrogate
\begin{equation}
\ell_{\mathrm{pay}}(y)
\;\triangleq\;
\ln\!\left(1+\beta\,|y|^2\right),
\label{eq:practical_bits_per_symbol}
\end{equation}
where $\beta$ summarizes the effective resolution per unit energy (and, implicitly, the target reliability level).

While actual semantic symbols are continuous, this logarithmic energy cost serves as a continuous relaxation of the discrete description length required to distinguish symbols. Specifically, as the minimum required distance $d_0$ between distinguishable symbols decreases, the number of discrete states $M(r)$ in \eqref{eq:Mr_affine} grows large, allowing the step-like counting function to be accurately approximated by the continuous payload surrogate $\ell_{\mathrm{pay}}(y)$. 
Notably, the form in \eqref{eq:practical_bits_per_symbol} aligns with Shannon’s $\log(1+\mathrm{SNR})$ law, interpreted here at an instantaneous level: higher symbol energy supports a larger local SNR within the symbol space, and consequently, accommodates a larger information payload.

Finally, we define the description-efficiency surrogate as the expected payload
\begin{equation}
\mathcal{L}_{\mathrm{desc}}
\;\triangleq\;
\mathbb{E}\!\left[\ell_{\mathrm{pay}}(Y)\right].
\label{eq:L_desc_def}
\end{equation}
This surrogate is not claimed to equal an exact codelength; it is a tractable proxy that captures the mechanism of implicit rate allocation through symbol energy in fixed-length continuous signaling.

\subsection{Joint Objective and Maximum-Entropy Solution}
\label{sec:joint_objective}

Recall that our symbol-shaping objective is posed directly over the input distribution $p(y)$ as
\begin{equation}
\underset{p(y)}{\text{minimize}}\quad
\mathcal{L}_{\mathrm{desc}}(p) + \lambda\,\mathcal{L}_{\mathrm{util}}(p),
\label{eq:lagrangian_objective_conceptual}
\end{equation}
where $\mathcal{L}_{\mathrm{util}}(p)\triangleq -h(Y)$ promotes channel utilization.
The description-efficiency term is defined via the energy-payload surrogate:
\begin{equation}
\mathcal{L}_{\mathrm{desc}}(p)\triangleq \mathbb{E}\!\left[\ell_{\mathrm{pay}}(Y)\right].
\label{eq:Ldesc_def_recall}
\end{equation}

\paragraph{From a 2D (I/Q) payload view to a scalar shaping model}
The APSK-inspired payload view is inherently two-dimensional: distinguishable states scale with the I/Q disk area.
In practical semantic encoders, however, the network produces real-valued features; two consecutive features are grouped to form one complex channel symbol.
Let $Y_{\mathbb{C}}\in\mathbb{C}$ denote a complex channel symbol formed by pairing two real encoder features, $Y_{\mathbb{C}}=Y_1+jY_2$.
For fitting and analysis, we model the common one-dimensional marginal of a real component $Y\in\{\Re\{Y_{\mathbb{C}}\},\Im\{Y_{\mathbb{C}}\}\}$ and use the scalar energy-payload surrogate
\begin{equation}
\ell_{\mathrm{pay}}(y)\triangleq \ln\!\left(1+\beta y^2\right).
\label{eq:Ldesc_ellpay_recall}
\end{equation}

\paragraph{Maximum-entropy form}
For any $\lambda>0$, minimizing \eqref{eq:lagrangian_objective_conceptual} is the unconstrained (Lagrangian) form of a maximum-entropy problem: there exists a constant $L_0$ (dependent on $\lambda$) such that the optimizer also solves
\begin{equation}
\begin{aligned}
\underset{p(y)}{\text{maximize}}\quad & h(Y)\\
\text{subject to}\quad 
& \mathbb{E}\!\left[\ell_{\mathrm{pay}}(Y)\right]=L_0, \int p(y)\,dy=1 .
\end{aligned}
\label{eq:max_entropy_problem}
\end{equation}
This is the standard KKT/Lagrangian correspondence: $\lambda$ selects the operating point $L_0$ on the trade-off curve.

\begin{proposition}
\label{prop:student_t_max_entropy}
The distribution that maximizes $h(Y)$ subject to $\mathbb{E}[\ell_{\mathrm{pay}}(Y)]=L_0$ is of the form
\begin{equation}
p(y)\propto \left(1+\beta y^2\right)^{-a},
\label{eq:t_kernel}
\end{equation}
which is a scaled Student's $t$ family.
\end{proposition}

\begin{proof}[Proof]
We maximize $h(Y)=-\int p(y)\ln p(y)\,dy$ subject to $\int p(y)\,dy=1$ and
\begin{equation}
\int p(y)\,\ln(1+\beta y^2)\,dy=L_0 .
\end{equation}
The Lagrangian functional is
\begin{equation}
\mathcal{J}[p]
=
\int \Big(
-p(y)\ln p(y)
-\mu_0 p(y)
-\mu_1 p(y)\ln(1+\beta y^2)
\Big)\,dy .
\end{equation}
Setting the functional derivative to zero yields
\begin{equation}
-\ln p(y)-1-\mu_0-\mu_1\ln(1+\beta y^2)=0,
\end{equation}
and therefore
\begin{equation}
p(y)
\propto
\exp\!\Big(-\mu_1 \ln(1+\beta y^2)\Big)
=
\left(1+\beta y^2\right)^{-\mu_1}.
\end{equation}
Defining $a\triangleq \mu_1>0$ gives \eqref{eq:t_kernel}. Rewriting with $s^2\triangleq 1/\beta$ yields
$p(y)\propto (1+y^2/s^2)^{-a}$, which is the kernel of a (scaled) Student's $t$-distribution.
\end{proof}

\paragraph{Variance-normalized one-parameter model}
The maximum-entropy solution \eqref{eq:t_kernel} has the kernel
$p(y)\propto (1+\beta y^2)^{-a}$ with two free parameters $(a,\beta)$, where $\beta$ is a scale-related parameter.
In practical semantic systems, a deterministic power-normalization layer enforces unit average power per real component, i.e., $\mathbb{E}[Y^2]=1$,
so the scale is not an independent degree of freedom.
Accordingly, we adopt the variance-normalized Student's $t$ family parameterized only by the degrees of freedom $\nu>2$:
\begin{equation}
p(y;\nu)=
\frac{\Gamma\!\left(\frac{\nu+1}{2}\right)}
{\sqrt{\pi(\nu-2)}\,\Gamma\!\left(\frac{\nu}{2}\right)}
\left(1+\frac{y^2}{\nu-2}\right)^{-\frac{\nu+1}{2}},
\label{eq:student_t_pdf}
\end{equation}
which satisfies $\mathbb{E}[Y]=0$ and $\mathbb{E}[Y^2]=1$.
With this normalization, the kernel parameters correspond to
$a=(\nu+1)/2$ and $\beta=1/(\nu-2)$.

\subsection{Implications of Coding Schemes and Training Dataset on Symbol Distributions}
\label{sec:implications}

The parameter $\nu$ in our model quantifies the inherent trade-off between the source objective (heavy-tailed) and the channel objective (Gaussian).
\rev{Smaller $\nu$ yields heavier tails, reflecting stronger reliance on energy-based implicit rate adaptation analogous to variable-length coding, whereas larger $\nu$ approaches a Gaussian-like, higher-entropy distribution, which is typically aligned with channel utilization over AWGN.}
Consequently, we expect larger $\nu$ values in scenarios where implicit power adaptation is less critical, such as with datasets having uniform entropy or architectures enabling explicit symbol-rate control.
Motivated by this view, we analyze end-to-end trained semantic communication systems focusing on two factors:

\textbf{Coding schemes.}
Architectures with fixed symbol budgets (e.g., DeepJSCC~\cite{bourtsoulatze2019deep}) have limited explicit rate control and are expected to compensate via energy-based adaptation,
potentially yielding heavier-tailed marginals (smaller $\nu$).
Conversely, architectures that explicitly vary the symbol budget (e.g., NTSCC~\cite{wang2023improved}) may reduce the burden on energy-based adaptation,
shifting the marginals toward Gaussian-like behavior (larger $\nu$).

\textbf{Training datasets.}
We further expect the rate variability of the training data to affect $\nu$.
When a dataset exhibits high entropy variability, a fixed-length transmitter must adapt more aggressively across samples, which can push the symbol marginals toward heavier tails (smaller $\nu$).
Conversely, more homogeneous datasets require less adaptation and tend to yield more Gaussian-like marginals (larger $\nu$).

\textbf{Training SNR.}
We further expect the training SNR to influence the learned tail parameter $\nu$.
At higher SNR, the model can use a smaller spacing (smaller $A_0$), which increases $\beta$ in \eqref{eq:practical_bits_per_symbol} (roughly, $\beta \propto 1/A_0$) and pushes the marginals toward heavier tails (smaller $\nu$).
Conversely, at lower SNR, the learned distribution shifts toward more Gaussian-like signaling (larger $\nu$).

In the next section, we empirically test these hypotheses by fitting $\nu$ from the learned symbol marginals across architectures and datasets.

\section{Empirical Validation of Symbol Distributions}
\label{sec:empirical_valid}
In this section, we empirically test whether the learned semantic symbols are well modeled by the proposed Student’s $t$ family and how the fitted shape parameter $\nu$ varies across architectures, datasets, and operating conditions.
Since the encoder does not provide an explicit density, we first approximate the symbol distribution from empirical samples and then fit $\nu$ to the resulting marginals.

\begin{figure}[t]
  \centering
  \includegraphics[width=0.9\columnwidth]{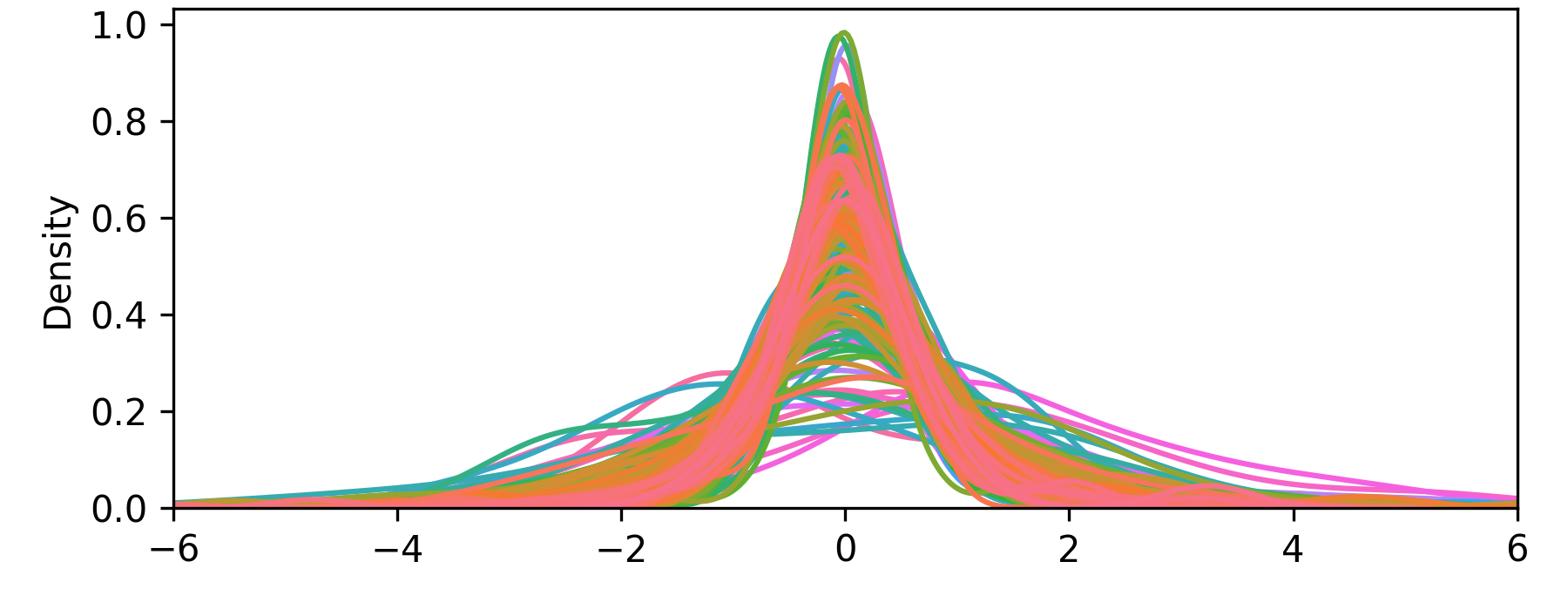}
  \caption{Empirical symbol distributions $q_i(y)$ for each symbol dimension, represented with distinct colors. The results are from a fixed-length model trained on the ImageNet dataset. The visualization illustrates that the empirical distributions across various dimensions exhibit similar zero-mean, bell-shaped distributions.}
  \label{fig:empirical_distribution}
\end{figure}

\subsection{Approximation of Symbol Distributions from Empirical Samples}
\label{sec:kde_approximation}

A key challenge is that the encoder output is deterministic, so the model does not provide an explicit likelihood.
Nevertheless, randomness of the input $X\sim q(\mathbf{x})$ induces a distribution over symbols.
Formally, with $\mathbf{y}=f_{\text{enc}}(\mathbf{x})$, the conditional distribution is a Dirac delta
\begin{equation}
q(\mathbf{y}\mid \mathbf{x})=\delta\!\left(\mathbf{y}-f_{\text{enc}}(\mathbf{x})\right),
\end{equation}
and the induced marginal is
\begin{equation}
q(\mathbf{y})=\int q(\mathbf{y}\mid \mathbf{x})\,q(\mathbf{x})\,d\mathbf{x}.
\end{equation}
Since $q(\mathbf{y})$ is not available in closed form, we estimate it from empirical samples.

Let $\{\mathbf{x}^{(b)}\}_{b=1}^{B}$ be a batch of inputs and $\mathbf{y}^{(b)}=f_{\text{enc}}(\mathbf{x}^{(b)})$ the corresponding deterministic symbol vectors.
Denote by $y_i^{(b)}$ the $i$-th component of $\mathbf{y}^{(b)}$.
We estimate the 1D marginal density of dimension $i$ via Gaussian-kernel density estimation (KDE):
\begin{equation}
\hat q_i(y)
\triangleq
\frac{1}{B}\sum_{b=1}^{B}
\mathcal{N}\!\left(y;\, y_i^{(b)},\, \sigma_{\text{KDE},i}^{2}\right),
\label{eq:kde_qi}
\end{equation}
where $\sigma_{\text{KDE},i}$ is the bandwidth. We set $\sigma_{\text{KDE},i}$ using Silverman's rule of thumb~\cite{silverman2018density}:
\begin{equation}
\sigma_{\text{KDE},i} \approx 1.06\, \hat{\sigma}_{i}\, B^{-1/5},
\end{equation}
with $\hat{\sigma}_{i}$ being the empirical standard deviation of $\{y_i^{(b)}\}_{b=1}^B$.

For tractability, we focus on a representative 1D marginal by approximating the marginal shapes as identical across dimensions, i.e., $q(y_i)\approx q(y)$ for all $i$.
This approximation is adopted to regulate the marginal distributional shape of individual symbol dimensions in a tractable manner.

Accordingly, we construct a pooled KDE as a common marginal:
\begin{equation}
\hat q(y)
\triangleq
\frac{1}{MB}\sum_{i=1}^{M}\sum_{b=1}^{B}
\mathcal{N}\!\left(y;\, y_i^{(b)},\, \sigma_{\text{KDE}}^{2}\right),
\label{eq:kde_q_pooled}
\end{equation}
where $\sigma_{\text{KDE}}$ is chosen using Silverman's rule applied to the pooled
samples (with effective sample size $MB$).

\begin{figure*}[t]
  \centering
  \includegraphics[width=0.9\textwidth]{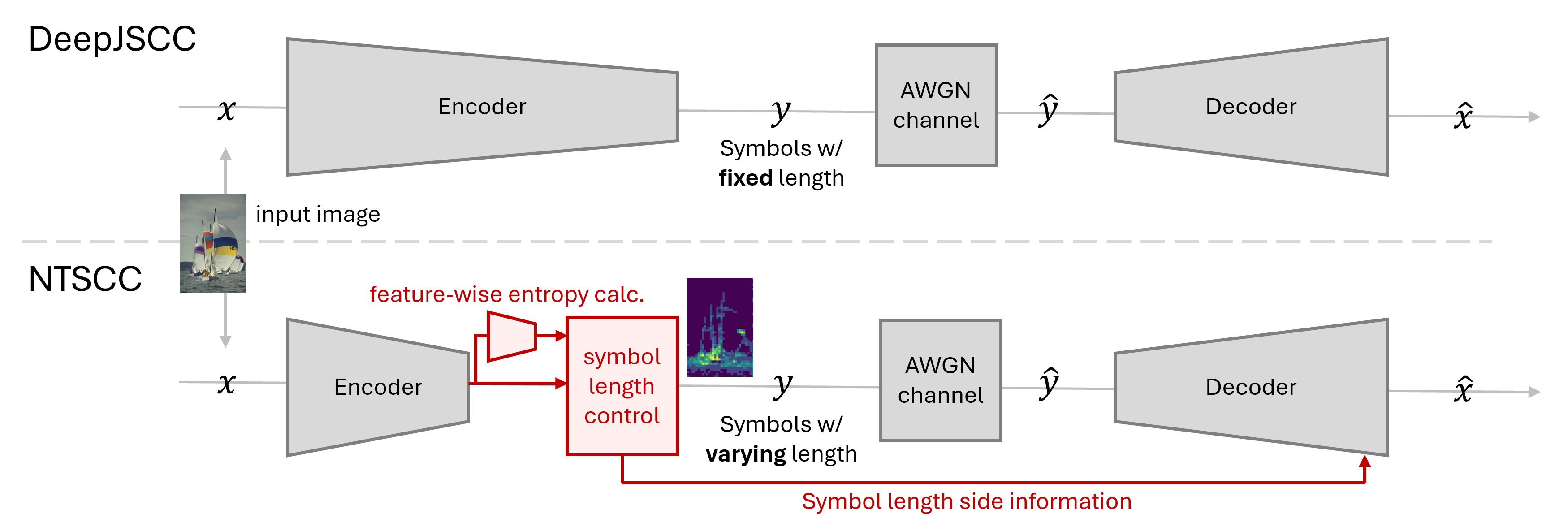}
  \caption{System architecture of DeepJSCC and NTSCC. The key difference in NTSCC compared to DeepJSCC is its feature-wise entropy calculation block, which enables explicit control over symbol length. This allows symbols to vary in length based on the image or extracted features, reducing the need for power-based variable-length coding.}
  \label{fig:deepjscc_ntscc_arch}
\end{figure*}

Using this common marginal, we employ the standard product-form approximation
\begin{equation}
\hat q(\mathbf{y}) \approx \prod_{i=1}^{M} \hat q(y_i),
\label{eq:kde_product_q}
\end{equation}
which serves as a tractable surrogate for distribution-level computations (e.g., visualization and regularization), rather than a statement of exact factorization.

This approximation is supported by (i) the impracticality of reliable per-dimension density estimation with small batches in large-scale datasets, and (ii) empirical evidence that different symbol dimensions exhibit similar zero-mean, bell-shaped marginals (see Fig.~\ref{fig:empirical_distribution}).
Furthermore, CIFAR-10 experiments in Fig.~\ref{fig:cifar_training_curve} indicate that using the pooled marginal has negligible impact on performance.

Note that the KDE is used only for qualitative visualization and for computing distribution-level regularizers.
When used for regularization, we approximate KL terms via Monte Carlo evaluation on encoder samples using the KDE densities (see Section~\ref{sec:proposed_method}).
For parametric fitting of the Student's $t$ model, we apply maximum likelihood estimation (MLE) directly to the pooled empirical samples $\{y_i^{(b)}\}_{i,b}$ (after variance normalization), without requiring KDE.


\subsection{Experimental Setup}
For our experiments, we consider an image semantic communication system designed to reconstruct images with minimal MSE. We evaluate the system across various image datasets, including CIFAR-10~\cite{cifar} and ImageNet~\cite{deng2009imagenet}, and explore different types of neural network architectures, such as ViT-based~\cite{yoo2023role} and CNN-based~\cite{he2022elic} models. We also examine different types of compression, comparing constant symbol rate~\cite{yoo2023role,he2022elic} with variable-length coding~\cite{wang2023improved}, and assess performance under various transmission rates (in symbols per pixel) and AWGN channel SNRs ranging from 0 to 20~dB.

\rev{The neural networks were trained with Adam using a learning rate of $10^{-4}$, with batch sizes of 8 for ImageNet and 32 for CIFAR-10. For ImageNet, we followed~\cite{he2022elic} and trained on randomly cropped \(256\times256\) patches from 8{,}000 images. Unless otherwise noted, we used the smaller model in~\cite{yoo2023role} for CIFAR-10 and the larger CNN-based model in~\cite{he2022elic} elsewhere. These models are constant-rate, meaning they generate a fixed number of symbols for each input image, unlike the variable-length coding used in NTSCC~\cite{wang2023improved}, which we will compare in Section~\ref{sec:coding_scheme}.}

For quantitative analysis, we fit the empirical symbols to our proposed Student's $t$-based symbol probability model and examine the fitted \(\nu\) to determine whether the symbol distribution is closer to a Gaussian or Cauchy distribution.  
Prior to fitting the unit-variance Student's $t$-model in \eqref{eq:student_t_pdf}, we normalize the symbol variance so that the only free parameter is the tail index $\nu$.
Specifically, to estimate the optimal \(\nu\) from empirical samples, we use MLE as follows:

\begin{equation}
\nu = \arg\max_{\nu} \sum_{i=1}^{M} \sum_{b=1}^{B} \log p(\mathbf{y}_{i}^{(b)}; \nu),
\end{equation}
where \( p(y; \nu) \) is the probability density function of the Student's $t$-distribution, as given in \eqref{eq:student_t_pdf}. Here, \( \mathbf{y}_{i}^{(b)} \) represents the \( i \)-th dimensional element of the symbol \( \mathbf{y} \) produced from the \( b \)-th image, \( M \) denotes the total number of dimensions, and \( B \) denotes the total number of images.
The estimated $\nu$ serves as a tail-index indicator: larger $\nu$ approaches the
Gaussian limit, while smaller $\nu$ yields heavier-tailed, Cauchy-like behavior.
Since the symbol variance is only defined for \( \nu > 2 \), we constrain the search space to \( (2, \infty) \). 

\begin{figure*}[htbp]
  \centering
  \includegraphics[width=0.93\textwidth]{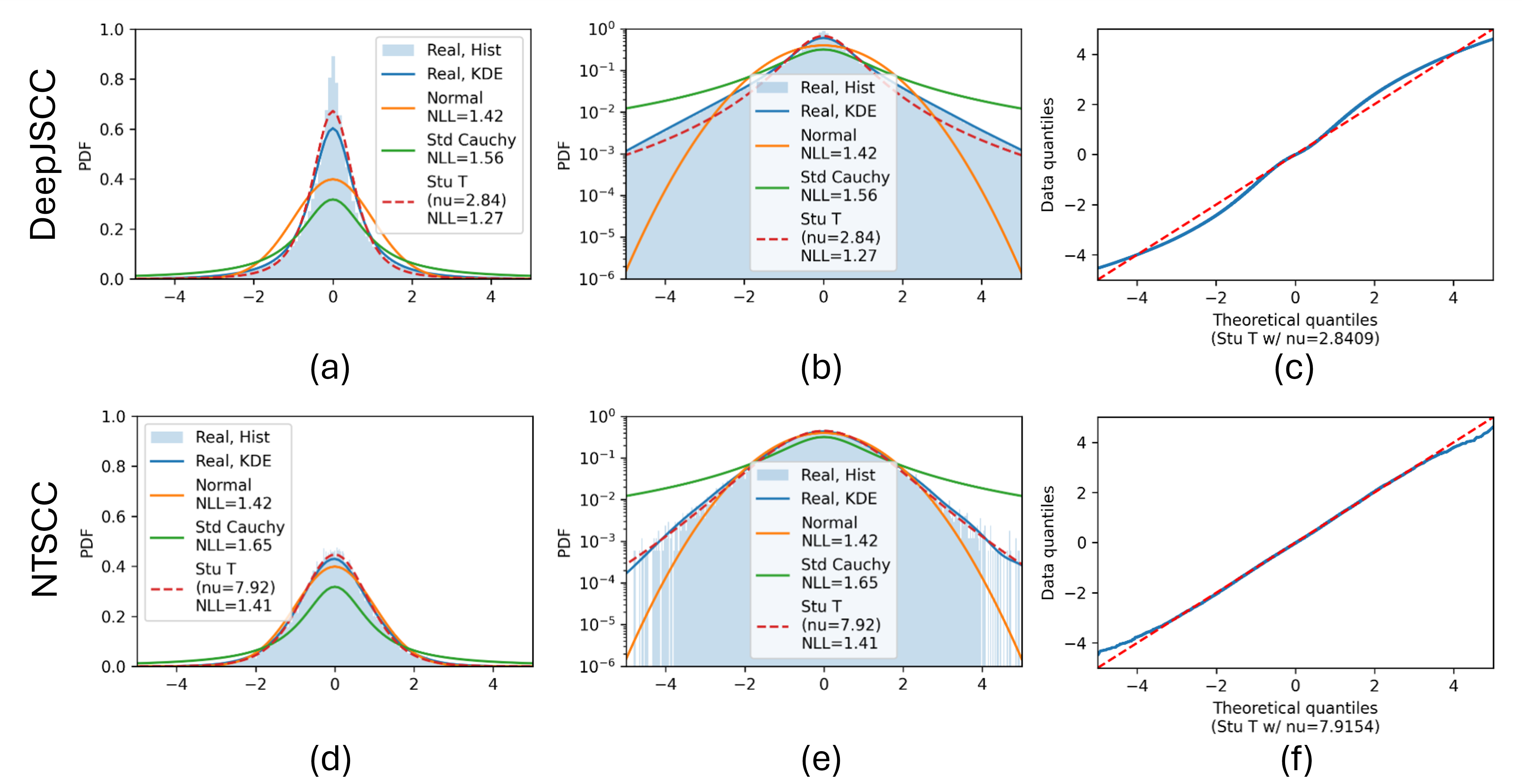}
  \caption{Symbol distributions with respect to coding schemes.
  (a, d) Linear-scale probability density function plots comparing the empirical symbol distributions (histogram and KDE) with Gaussian, Standard Cauchy, and Student's $t$-distributions. 
  (b, e) Log-scale PDF plots highlighting differences in the tail behavior of each distribution.  
  (c, f) Quantile-Quantile plots comparing the empirical symbol distributions to the fitted Student's $t$-model, assessing the goodness of fit.
  The top row represents the results from DeepJSCC model, while the bottom row corresponds to NTSCC model.
  }
  \label{fig:symbol_distributions_compression_type}
\end{figure*}

\subsection{Symbol Distributions with Respect to Coding Schemes}
\label{sec:coding_scheme}

\rev{To validate our hypothesis, we compare the symbol distributions of DeepJSCC~\cite{bourtsoulatze2019deep} and NTSCC~\cite{wang2023improved}. DeepJSCC uses a fixed symbol budget, whereas NTSCC explicitly adapts the number of transmitted symbols to image entropy (Fig.~\ref{fig:deepjscc_ntscc_arch}), reducing reliance on power-based rate allocation.}

As hypothesized in the previous section, a smaller $\nu$ (heavier-tailed, Cauchy-like tail behavior) indicates stronger reliance on power-based implicit rate adaptation, which plays a role analogous to variable-length coding in fixed-length continuous signaling. In contrast, Gaussian-like symbols (large $\nu$) are aligned with channel utilization, approaching Gaussian-like (high-entropy) signaling under a unit power constraint. Consequently, DeepJSCC, which uses a fixed symbol budget and lacks explicit rate control, is expected to learn symbol marginals between the Gaussian and heavy-tailed regimes. By contrast, NTSCC explicitly varies the number of transmitted symbols, reducing the burden on power-domain adaptation and pushing the learned distribution closer to Gaussian.

\begin{figure*}[htbp]
  \centering
  \includegraphics[width=0.93\textwidth]{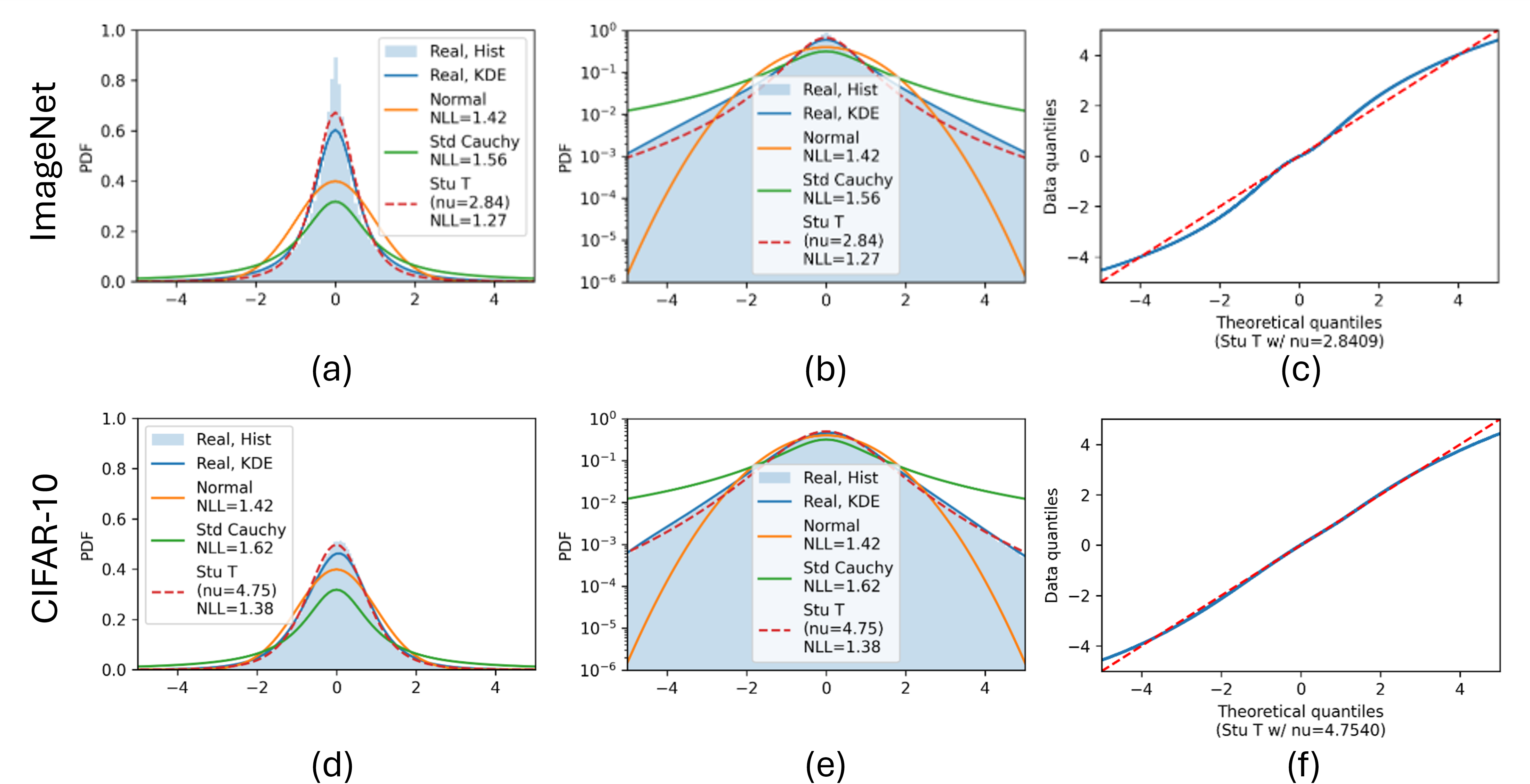}
  \caption{Symbol distributions with respect to training datasets.  
  (a, d) Linear-scale probability density function plots comparing the empirical symbol distributions (histogram and KDE) with Gaussian, Standard Cauchy, and Student's $t$-distributions.  
  (b, e) Log-scale PDF plots highlighting differences in the tail behavior of each distribution.  
  (c, f) Quantile-Quantile plots comparing the empirical symbol distributions to the fitted Student's $t$-model, assessing the goodness of fit.
  The top row represents ImageNet-trained models, while the bottom row corresponds to CIFAR-10-trained models.
  }
  \label{fig:symbol_distributions_datasets}
\end{figure*}

Fig.~\ref{fig:symbol_distributions_compression_type} illustrates the symbol distributions of these two systems.
Here, we provide histograms of symbols, KDE-estimated symbol distributions, a Gaussian distribution, a Standard Cauchy distribution, and a Student's $t$-distribution with a fitted \(\nu\). As anticipated, the results show that DeepJSCC produces distributions that exhibit heavier tails towards a Cauchy-like shape (fitted $\nu=2.84$). Meanwhile, NTSCC demonstrates distributions that align more closely with a Gaussian distribution ($\nu=7.92$).
The average NLL was 1.27 for DeepJSCC and 1.41 for NTSCC, which indicates a reasonable fit and is lower than that of naive Gaussian or standard Student's $t$-based modeling. 
Note that lower NLL values correspond to higher likelihoods.  

While the tail distribution appears to be reasonably well captured by our model, some deviations are observed in sparsely sampled regions, particularly at higher quantiles.  
These discrepancies may stem from the limited availability of symbol samples in those regions or the representational limitations of our Student's $t$-model.
Nevertheless, these empirical observations broadly support our model's predictions, illustrating the influence of the chosen coding scheme on symbol distributions in semantic communication systems.

\subsection{Symbol Distributions with Respect to Training Datasets}

\begin{figure}[t]
  \centering
  \includegraphics[width=0.85\columnwidth]{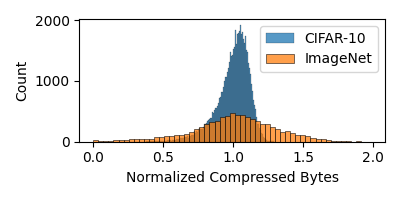}
  \caption{Compressed rate distribution of the randomly cropped input images from ImageNet (8,000 images) and CIFAR-10 datasets (50,000 images).}
  \label{fig:image_rates}
\end{figure}

We investigate how the variability in dataset entropy influences symbol distributions. The entropy variance among samples in a dataset directly affects the effectiveness of variable-length coding, subsequently impacting the resulting symbol distribution. To approximate the entropy variability, we compare the losslessly compressed sizes of images (using the PNG format) across datasets.
As shown in Fig.~\ref{fig:image_rates}, the ImageNet dataset exhibits significantly higher variability in image entropy compared to CIFAR-10. This increased entropy variance makes variable-length coding more advantageous for ImageNet, as it allows the system to adapt symbol lengths dynamically to match varying information content across images.

The effect of this entropy variability is reflected in the resulting symbol distributions. Fig.~\ref{fig:symbol_distributions_datasets} shows that the symbol distributions for ImageNet-trained systems lean more towards a Cauchy-like heavy tailed distribution ($\nu=2.84$), while CIFAR-10-trained systems exhibit distributions closer to Gaussian ($\nu=4.75$). This observation aligns with our predictions: datasets with higher entropy variance favor variable-length coding, naturally driving the symbol distribution toward a heavy tailed region. Conversely, datasets with relatively uniform entropy, such as CIFAR-10, reduce the need for variable-length coding, resulting in distributions that more closely approximate a Gaussian distribution. The NLL was 1.27 for ImageNet and 1.38 for CIFAR-10, respectively. These results emphasize the interplay between dataset characteristics and symbol distribution shaping, highlighting the adaptive nature of neural networks in balancing information maximization and bit-description cost minimization based on training data properties.

\subsection{Symbol Distributions with Respect to Channel SNR}
We further examine the impact of channel SNR on the learned symbol distributions.
To this end, we evaluated the symbol distributions of our semantic communication system under different SNR conditions by fitting the empirical samples to our proposed Student's $t$-based model.
The following SNR values were considered: 20~dB, 10~dB, and 0~dB.
Unless otherwise noted, the encoder/decoder are retrained for each SNR. 
Thus, the observed change in $\nu$ reflects how training adapts the output distribution under different noise levels.

\begin{figure*}[htbp]
  \centering
  \includegraphics[width=0.93\textwidth]{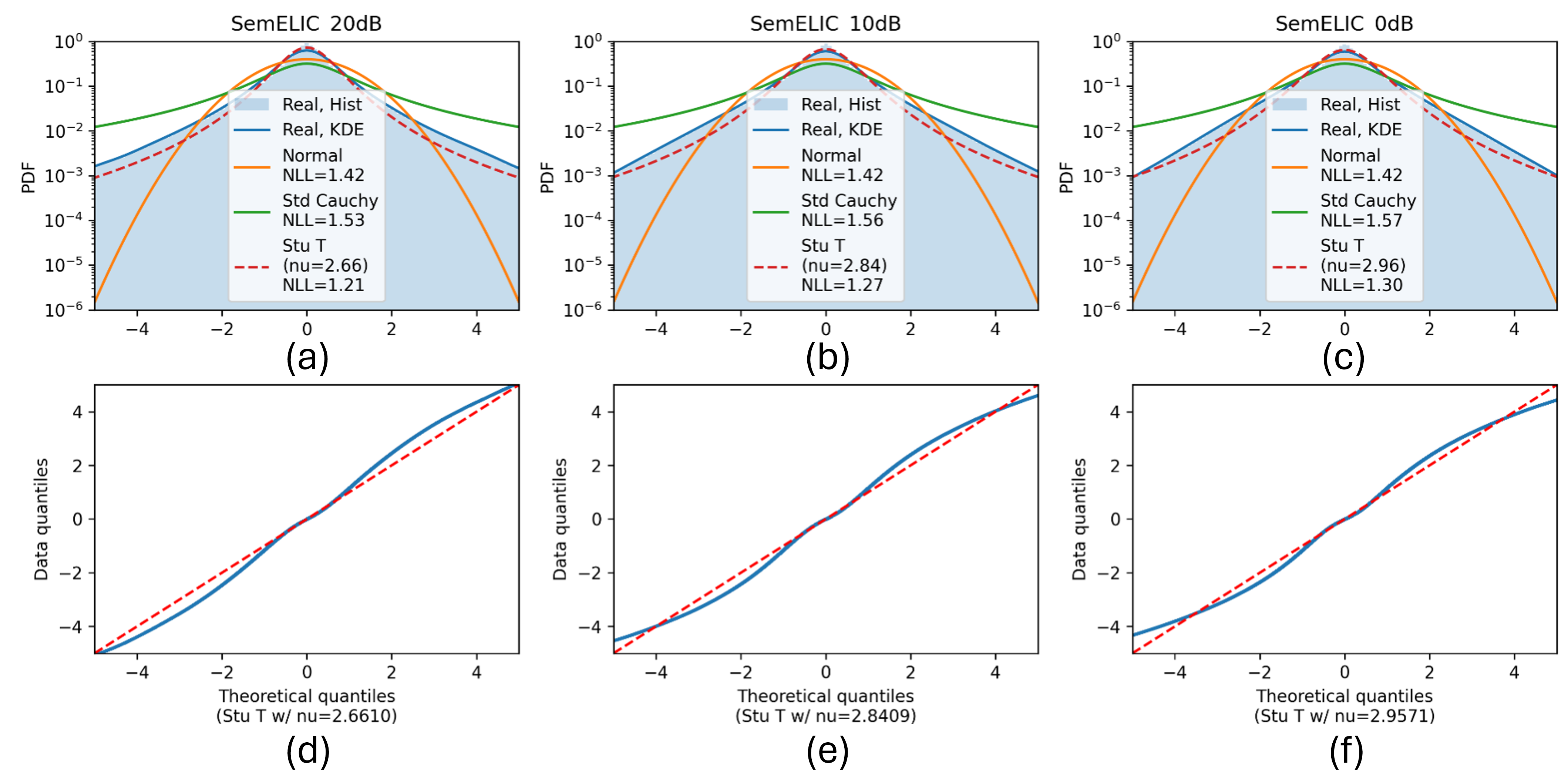}
  \caption{Symbol distributions with respect to channel SNRs.  
  (a, b, c) Log-scale probability density function plots comparing the empirical symbol distributions (histogram and KDE) with Gaussian, Standard Cauchy, and Student's $t$-distributions.  
  (d, e, f) Quantile-Quantile plots comparing the empirical symbol distributions to the fitted Student's $t$-model, assessing the goodness of fit.  
  The columns correspond to different SNR conditions: 20~dB (left), 10~dB (middle), and 0~dB (right).
  }
  \label{fig:symbol_distributions_SNR}
\end{figure*}

Fig.~\ref{fig:symbol_distributions_SNR} presents the results.
At an SNR of 20~dB, the fitted \(\nu\) is 2.66 with an average NLL of 1.21.
At 10~dB, \(\nu\) increases to 2.84 (NLL 1.27), and at 0~dB it further increases to 2.96 (NLL 1.30).
Although the absolute change in \(\nu\) is modest, the trend is consistent across SNRs: higher SNR leads to smaller \(\nu\), i.e., heavier-tailed symbol marginals.


This trend is also consistent with the proposed energy-payload surrogate.
At higher SNR, reliable discrimination requires a smaller decision region area \(A_0\), increasing the number of distinguishable states per unit energy.
In \eqref{eq:practical_bits_per_symbol}, this corresponds to a larger effective \(\beta\) (with \(\beta \propto 1/A_0\)), which strengthens the benefit of energy-based rate adaptation and favors heavier-tailed distributions (smaller \(\nu\)).
Conversely, at lower SNR, reliable discrimination is more limited (effectively larger \(A_0\) and smaller \(\beta\)), reducing the gain from power-based adaptation and shifting the learned distribution toward more Gaussian-like signaling (larger \(\nu\)), where channel utilization becomes more critical.

\section{Proposed Distribution-Regulating Loss Function}
\label{sec:proposed_method}
\rev{To further examine the practical relevance of our framework, we test whether explicitly guiding the symbol distribution affects system training. This is used primarily as an empirical probe of the proposed source-channel interpretation, rather than as a claim of a universally superior training objective.}

\subsection{Proposed Loss Function}
The proposed loss is defined as
\begin{equation}
\mathcal{L}(\theta)
= d(\mathbf{x}, \hat{\mathbf{x}})
+ \lambda \, \KL\!\left(\hat q(\mathbf{y}) \,\|\, p(\mathbf{y})\right),
\label{eq:loss_kl}
\end{equation}
where $d(\cdot,\cdot)$ is a task distortion metric (e.g., MSE or cross-entropy) and
$p(\mathbf{y})$ is a chosen target prior (e.g., a zero-mean unit-variance Gaussian).
Since the deterministic encoder does not provide an explicit likelihood, we approximate the encoder-output distribution by the pooled KDE-based product-form surrogate $\hat q(\mathbf{y})$ in \eqref{eq:kde_product_q}.

The KL term penalizes the mismatch between the empirical symbol distribution and the target prior,
\begin{equation}
\KL\!\left(\hat q(\mathbf{y}) \,\|\, p(\mathbf{y})\right)
\triangleq
\mathbb{E}_{\mathbf{y}\sim \hat q}\!\left[\log \hat q(\mathbf{y}) - \log p(\mathbf{y})\right].
\label{eq:kl_def_hatq}
\end{equation}
In practice, we approximate \eqref{eq:kl_def_hatq} by Monte Carlo averaging over encoder outputs
$\mathbf{y}^{(b)} = f_{\mathrm{enc}}(\mathbf{x}^{(b)})$:
\begin{equation}
\KL\!\left(\hat q \,\|\, p\right)
\approx
\frac{1}{B}\sum_{b=1}^{B}\Big(\log \hat q(\mathbf{y}^{(b)}) - \log p(\mathbf{y}^{(b)})\Big),
\label{eq:kl_mc}
\end{equation}
where $\log \hat q(\mathbf{y}^{(b)})$ is evaluated using the KDE density estimate.
This regularization encourages the encoder to produce symbol marginals that match the desired prior while preserving task performance through the distortion term.

While our analysis suggests that a Student's~$t$ distribution may be a reasonable prior family, its appropriate degrees-of-freedom parameter $\nu$ is not known a priori and would require re-estimation (or tuning) during training to evaluate the KL term, which can be unstable and computationally costly. 
We therefore adopt a zero-mean unit-variance Gaussian prior for $p(\mathbf{y})$ as a simple and stable choice. 
This corresponds to the limiting case $\nu\to\infty$ of the Student's~$t$ family and reduces the hyperparameter search to the regularization strength $\lambda$ only.
\rev{The proposed KDE-based regularizer adds computation only during training. Its cost comes from KDE density evaluation over the analyzed symbol samples, which scales quadratically with the number of samples used in the KDE computation, while leaving the network architecture, parameter count, and inference-time complexity unchanged. In our CIFAR-10 setting (512 symbols, 10~dB), this corresponded to an approximately 4.8\% increase in wall-clock training time over 100 epochs. }

\subsection{Performance Evaluation across Compression Rate and Channel SNR}
We first evaluate the performance of the semantic communication system with the proposed loss on the CIFAR-10 dataset under varying compression ratios and channel SNR conditions. We report the channel bandwidth ratio (CBR) following \cite{bourtsoulatze2019deep, weng2021semantic}, where CBR is computed as
\begin{equation}
\text{CBR} = \frac{S}{H \times W \times C},
\end{equation}
where \(S\) represents the total number of complex-valued symbols per image and \(H \times W \times C\) corresponds to the input image dimensions.

\begin{figure*}[t]
  \centering
  \includegraphics[width=0.99\textwidth]{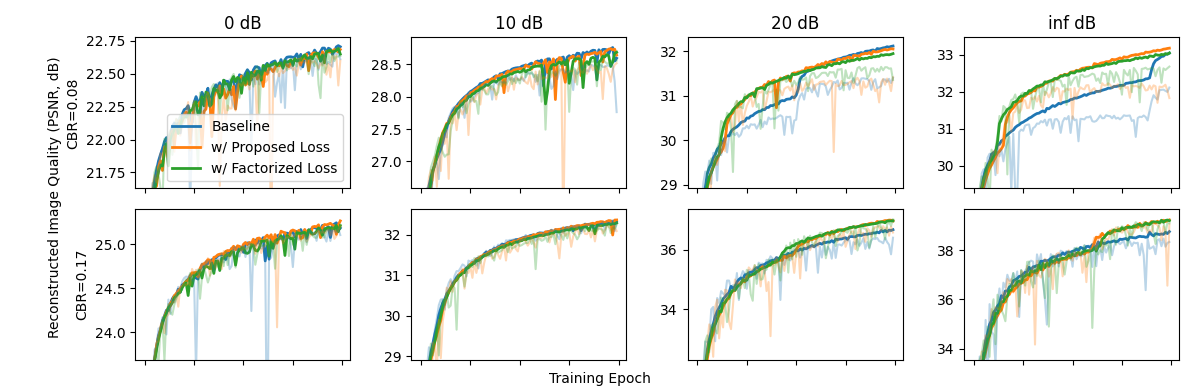}
  \caption{Training curves of the semantic system with (orange) and without the proposed loss (blue) on the CIFAR-10 dataset, evaluated with respect to compression ratio and channel SNRs.  
Curves with reduced opacity correspond to the validation set. The green curves represent the proposed loss without the identical-marginals assumption (see Section~\ref{sec:kde_approximation}).}
  \label{fig:cifar_training_curve}
\end{figure*}

\rev{The training curves in Fig.~\ref{fig:cifar_training_curve} illustrate the impact of our proposed loss function. A small regularization parameter of \(\lambda = 10^{-4}\) was used, as higher values degraded performance (not shown), while smaller values had negligible impact, behaving similarly to the original loss function without the KL term. The results indicate that our method primarily accelerates training, particularly in high-compression and high-SNR settings, where source coding considerations are more significant. This suggests that weakly regulating the symbol distribution toward a Gaussian distribution can improve convergence behavior.}

\rev{A possible explanation is that in source-coding-dominated regimes, where the symbol distribution tends to be more heavy-tailed, the model struggles more to learn a stable representation. A mild Gaussian regularization with low \(\lambda\) can therefore improve training stability and convergence.}

\rev{We also compare identical- and non-identical-KDE variants of the proposed loss (green curve in Fig.~\ref{fig:cifar_training_curve}). Their training behavior is similar overall, with the non-identical version showing slightly better validation performance. For larger datasets and models, however, we use the identical-KDE version because of batch-size and memory limitations.}

\begin{figure*}[!t]
  \centering
  \subfloat[]{%
      \includegraphics[width=0.31\textwidth]{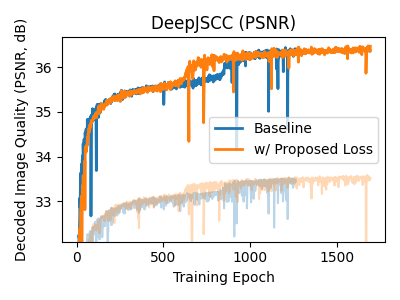}
      \label{fig:DeepJSCC_training_curves_PSNR}
  }
  \hfill
  \subfloat[]{%
      \includegraphics[width=0.31\textwidth]{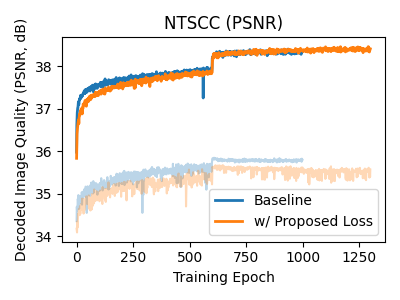}
      \label{fig:NTSCC_training_curves_PSNR}
  }
  \hfill
  \subfloat[]{%
      \includegraphics[width=0.31\textwidth]{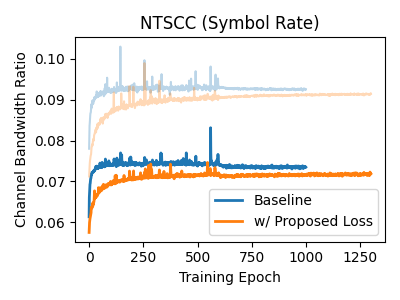}
      \label{fig:NTSCC_training_curves_CBR}
  }
  \caption{(a), (b) Training curves showing reconstructed image quality of the semantic system with and without the proposed loss for DeepJSCC and NTSCC, respectively. 
  The DeepJSCC model has channel bandwidth ratio of 0.083.
(c) Training curves depicting the average compression rate for NTSCC.  
SNR was set to 10~dB, and curves with reduced opacity correspond to the validation set.
}
\end{figure*}

\subsection{Performance Evaluation across System Architecture}
We then compare DeepJSCC and NTSCC with and without the proposed loss on ImageNet to isolate the impact of fixed-length (DeepJSCC) versus variable-length (NTSCC) transmission in a large-scale dataset.
The Kodak dataset~\cite{franzen1999kodak} is used for validation.
Since NTSCC transmits a variable number of symbols, we report both reconstruction quality and the achieved symbol rate (compression ratio), as higher rates naturally yield higher PSNR.
For reference, our DeepJSCC model operates at a channel bandwidth ratio of 0.083, i.e., 16{,}384 complex symbols per \(256\times256\) color image.

Fig.~\ref{fig:DeepJSCC_training_curves_PSNR}-\ref{fig:NTSCC_training_curves_CBR} illustrate the training dynamics of DeepJSCC and NTSCC under the baseline and the proposed loss.
Across the tested regularization strengths (\(\lambda\in[10^{-4},1]\)), we report the best-performing settings (\(\lambda=10^{-4}\) for DeepJSCC and \(\lambda=1\) for NTSCC).
For DeepJSCC, the observed trend is consistent with CIFAR-10, supporting the generality of our observations.
In contrast, NTSCC is largely insensitive to \(\lambda\): sweeping \(\lambda\) from \(10^{-4}\) to \(1\) yields nearly identical trajectories.
This behavior is consistent with our framework: NTSCC already provides explicit rate control and produces near-Gaussian symbols (fitted \(\nu=7.92\)), so additional Gaussian regularization is largely redundant.

\subsection{Discussions}
While our analysis provides a robust source-channel framework for the AWGN channel, it also opens several important directions for practical deployment and theoretical extension:

\begin{itemize}
\item \textbf{Peak-Power Limits and Analysis of Clipped Distributions.} Real-world transmitters impose strict peak-power limitations, which practically truncate the heavy tails of the observed Student's $t$-distribution. Future theoretical research should focus on analyzing and optimizing the source-channel balance for this clipped Student's $t$-distribution model.

\item \textbf{More Diverse Channels and Practical Constraints.} We focus on the AWGN channel and continuous-valued analog signaling because this setting isolates the proposed source-channel trade-off in the cleanest form. Extending the framework to fading, multiple-input multiple-output (MIMO), nonlinear, or digital/quantized systems, where channel uncertainty, spatial coupling, clipping, and discrete signaling constraints also matter, is an important direction for future work.

\end{itemize}

\section{Conclusion}
\label{sec:conclusion}
This paper analyzed learned symbol distributions in end-to-end semantic communication by linking the distortion objective to two induced pressures on symbol shaping: channel utilization and description efficiency.
Using tractable surrogate objectives for these pressures, we derived a Student's $t$ symbol model and interpreted its tail parameter as a single knob reflecting relative strength of each pressure.
Our model predicted well how the shape parameter varies with the system’s reliance on power-domain rate control and with the sample-to-sample entropy variability of the training dataset.
\rev{Furthermore, the proposed KL regularizer toward a Gaussian target improved training stability and often accelerated convergence, providing supporting empirical evidence for the framework.}
Overall, these insights move beyond a black-box view of neural encoders and provide a principled basis for designing more stable and efficient semantic communication systems.


\bibliographystyle{ieeetr}
\bibliography{ref}

\vfill

\end{document}